\def\year{2020}\relax
\documentclass[letterpaper]{article} % DO NOT CHANGE THIS
\usepackage{aaai20}  % DO NOT CHANGE THIS
\usepackage{times}  % DO NOT CHANGE THIS
\usepackage{soul}
\usepackage{color}
\usepackage{booktabs}

\usepackage{amsthm}
\usepackage{amsmath}
\usepackage{mathtools}
\usepackage{helvet} % DO NOT CHANGE THIS
\usepackage{courier}  % DO NOT CHANGE THIS
\usepackage[hyphens]{url}  % DO NOT CHANGE THIS
\usepackage{graphicx} % DO NOT CHANGE THIS
\usepackage{graphicx}  % DO NOT CHANGE THIS
\theoremstyle{definition}
\newtheorem{mydef}{Definition}
\newtheorem{example}{Example}
\mathchardef\mhyphen="2D

\title{On the Evaluation of Intelligent Process Automation}
\author{
Deborah Ferreira$^1$, Julia Rozanova$^1$, Krishna Dubba$^2$, Dell Zhang$^2$\thanks{Dell Zhang is on leave from Birkbeck, University of London.}, Andre Freitas$^1$\\
 $^1$Department of Computer Science, University of Manchester, Kilburn Building, Oxford Road, M13 9PL, UK \\
 $^2$Blue Prism AI Labs, c/o WeWork, 125 Kingsway, London WC2B 6NH, UK\\
 \{deborah.ferreira, julia.rozanova,  andre.freitas\}@manchester.ac.uk, \{krishna.dubba, dell.zhang\}@blueprism.com}

\begin{document}

\maketitle

\begin{abstract}

Intelligent Process Automation (IPA) is emerging as a sub-field of AI to support the automation of long-tail processes which requires the coordination of tasks across different systems. So far, the field of IPA has been largely driven by systems and use cases, lacking a more formal definition of the task and its assessment. This paper aims to address this gap by providing a formalisation of IPA and by proposing specific metrics to support the empirical evaluation of IPA systems. This work also compares and contrasts IPA against related tasks such as end-user programming and program synthesis.

\end{abstract}

\section{Introduction}

Robotic Process Automation (RPA) aims to provide a supporting framework to automate the long-tail of processes which involve routine tasks, structured data and deterministic outcomes~\cite{10.1007/978-3-319-66963-2_7}. RPA supports end-users in the automation of existing processes without the requirement of a programming language. 

Even though there has been an increase in investments in the area of RPA, it still relies on the explicit encoding of rules and configurations for process generation, with limited use of AI methods. Agostinelli et al.~\cite{agostinelli_marrella_mecella} provides an analysis of several RPA tools, none of the studied tools has self-learning ability and, are not able to automatically understand which actions belong to which process (\textit{intra-routine learning}) and which processes are available for automation (\textit{inter-routine learning}). 

The goal of Intelligent Process Automation (IPA) is to generalise RPA, proving the tools to create complex workflows, with minimal user interference~\cite{reddy2019}. IEEE Standards Association defines Intelligent Process Automation as ``a preconfigured software instance that combines business rules, experience based context determination logic, and decision criteria to initiate and execute multiple interrelated human and automated processes in a dynamic context.''~\cite{8070671}.

Consider, for example, a user that works in the IT department of a company and is responsible for redirecting tickets from the request system to the correct department. RPA would allow the user to automate this process by manually generating a set of rules capable of redirecting each ticket to the appropriate department; however, if the process is complex, with several variables, the implementation might be costly and slow. IPA systems, on the other hand, would observe the user's actions and detect the patterns between different requests and the redirected departments. With sufficient examples and the clarification of the user intents, it would automate the process, minimising human intervention.

So far, the field of IPA has been largely driven by systems and use cases, lacking a more formal definition of the task ans its systematic evaluation. This paper aims to address this gap by focusing on the following contributions:
\begin{enumerate}
    \item Providing a formalisation of IPA.
    \item Proposing specific metrics to support the empirical evaluation of IPA methods and systems.
    \item Introducing a new benchmark for the evaluation of IPA.
    \item Comparing and contrasting IPA against related tasks such as end-user programming and program synthesis.
\end{enumerate}

This paper is organised as follows: Section 2 presents the formalisation of IPA and related concepts. Section 3 introduces the modalities and tasks that are part of IPA. Section 4 analyses research areas that are similar to IPA. Section 5 then uses the formalism defined previously to define metrics for the different IPA tasks. Section 6 presents the methodology used for constructing a benchmark for evaluating IPA tasks. Finally, Section 7 concludes this work.

% \hl{\emph{Skeleton:}}

% Introduction

% What is IPA? 

% Motivational scenario

% Problem description and formalisation:

% Demo to process

% vid!

% Text to process

% Demo to text

% (We can communicate as a triangle between these different representation modalities)

% Discuss what we are trying to evaluate (in natural language)`	

% \hl{\emph{Draft:}}

% The automation of repetitive and menial processes with minimal direction is a key goal of artificial intelligence. Intelligent Process Automation is the goal of automatically producing an executable workflow out of a brief demonstration or description of a task to be executed in a digital environment. 

\section{Formalising IPA}

At the center of IPA is the capture and formalisation of a workflow (the interaction between end-user actions, software and data artefacts within an end-user interface environment) in a formal language which can be used to re-enact the workflow.

IPA formal languages have particular properties which aims at eliciting the end-user actions embedded within a workflow. The formalisation of these properties is a pre-requisite for the definition of an evaluation methodology. It is through this language that we may understand and evaluate a given \textit{interactor}, such as the Sikuli GUI automation tool\footnote{http://sikulix.com/}. Operating directly with the interactor allows us to evaluate the interaction of the end processes with the same interfaces a human would encounter, rather than the extremely varied computations that underlie the systems with which it interacts.
To this end, we propose an abstract \emph{interaction language} with which to describe the kind of interface-centred activity to be mimicked by IPA. We first define it purely syntactically, and then describe its intended instantiation in terms of GUIs and GUI interactions. 
\newline

\begin{mydef}
An \textbf{interface} is a finite set $I$ of symbols $\{i_1, \ldots, i_n\}$ which will be referred to as \emph{interface elements}. 
\end{mydef}

% \begin{mydef}
% An \textbf{environment} $e$ is defined by a $4$-tuple $(\mathcal{I}, F, \Omega, S)$ where $\mathcal{I} = {I_1, \ldots, I_n}$ is a finite set of interfaces, $\Omega$ is a set of values available to the end-user as potential inputs, $S$ is a set of computational states of a system and $\mathcal{F}$ is a set of interactor functions of the form $f: S 
% \times \bigcup \{\Omega, I_1, 
% \ldots, I_n \} \to S$ .
% \end{mydef}

% In practice, the computational state $s 
% \in S$ will be treated as a black box on which the interaction functions have an effect, but which we may not be able to describe or measure (nor do we wish to). 

%%%% LOOKING FOR THIS

% \begin{mydef}
% An \textbf{environment} $e$ is defined by a $4$-tuple $(\mathcal{I}, \mathcal{F}, \Omega, s)$ where $\mathcal{I} = {I_1, \ldots, I_n}$ is a finite set of \emph{interfaces}, $\Omega$ is a set of \emph{values} available to the end-user as potential inputs and $F$ is a set of finitary function symbols $f(s, arg_1, \ldots arg_m)$ which will represent available interface interactions, where each $arg$ may be any interface element $i 
% \in I_j$ or a value $v \in \Omega$. 

% Finally, the symbol $s$ is representative of the current computational state, which will always serve as the first argument to our function symbols. 
% \end{mydef}

\begin{mydef}
\label{def_env}
An \textbf{environment} $e$ is defined by a $4$-tuple $(\mathcal{I}, \mathcal{F}, \Omega,  S)$ where $\mathcal{I} = {I_1, \ldots, I_n}$ is a finite set of \emph{interfaces}, $\Omega$ is a countably infinite set of variables (\emph{$arg_1, arg_2,\ldots$}) or \emph{value symbols} and $F$ is a set of finitary function symbols $f(s, arg_1, \ldots arg_m)$ where each $arg$ may be any interface element $i \in I_j$ or a value $v \in \Omega$. $S$ will be referred to as the \emph{state space}.
\end{mydef}

The following example serves to introduce the intended concrete realisations of the described language, after which we turn to the definitions relevant to its interpretation.

\begin{example}
\begin{figure}[htbp!]
    \centering
    \resizebox{\columnwidth}{!}{
    \begin{tabular}{|c|c|}
    \hline
        &\\
        $ I_1 $ & $ I_2 $  \\ 
        \includegraphics[width=7cm, trim={1cm 13cm 17cm 1cm}, clip]{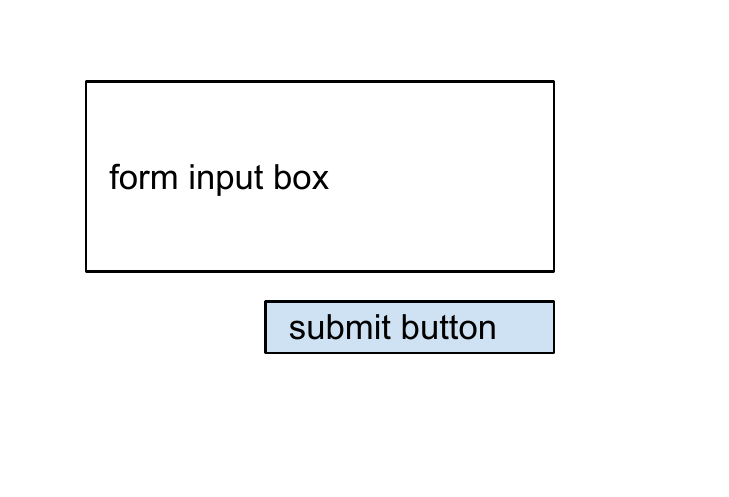}  & \includegraphics[width=7cm, trim={1cm 13cm 17cm 1cm}, clip]{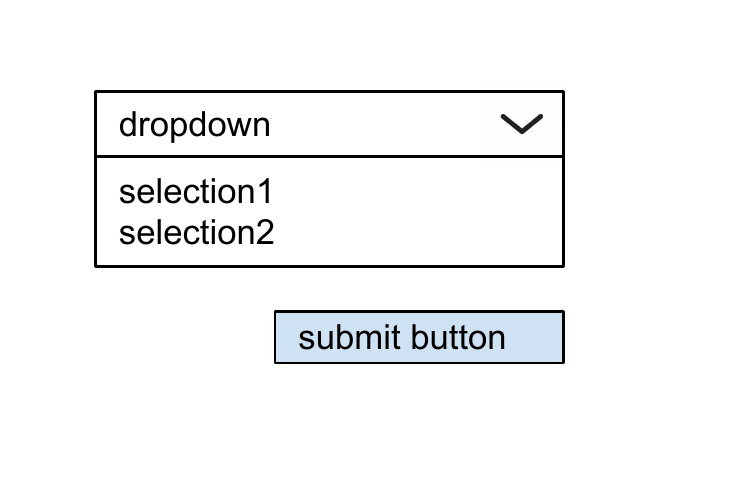} \\
        \hline
    \end{tabular}
    }
    \caption{A simplified environment $e$ with two interfaces, $I_1$ and $I_2$.}
    \label{two_interface_example}
\end{figure}

An end-user may work on a single desktop configuration with multiple windows, each of which can be identified as a separate interface. We may assume the interface elements are represented by unique IDs.

In Figure \ref{two_interface_example}, let $I_1 = \{i_{1}^{(1)}, i_{2}^{(1)} \}$, where these identify the form input box and the submit button respectively.  The state symbol $s$ will represent the computational state under the hood, while the function set $F$ is determined by the valid actions relevant to the interfaces: here, for example, one can imagine these to include a \emph{click} action $f_1(s, i)$ and a \emph{text input} action $f_2(s, i, ~\mbox{`example text input'})$. 

The set $\Omega$ is that of all possible user inputs to the text box. To simplify, suppose that in this case it is \mbox{$\Omega = \{`a\textrm',
\ldots,`z\textrm',`\ \ \textrm' \}^*$} (where $^*$ is the Kleene star).

\end{example}

% \noindent \textbf{Definition (Environment): } An environment $e$ is defined by a set $F$ of systems functions, where $f(args ...)$ and a set of systems' interfaces $I$. 

% Define the mapping between the interface and the args here. Define the type of an interface element i here.
% \newline

% \noindent \textit{Example:} Use a concrete textbox + button as an example. textbox maps to argument, button maps to the function.
% \newline

\begin{mydef}
A model or \emph{instantiation} of an environment $e$ is defined with respect to a working computer desktop with finitely many graphical user interfaces, an \emph{interactor} (which may either be automatic, such as a Sikuli program instance, or a human end-user) which has its own language and an interpretation function $[[\cdot]]$ which will be described comprehensively in the next definition.
\end{mydef}

\begin{mydef}{\textbf{Interpretation Function:}}

Given an environment $e = (\mathcal{I}, F, \Omega, S)$ and a fixed interactor language, we define the following interpretation function:

\begin{itemize}
    \item $[[S]]$ is abstractly defined to be the computational state space of the machine on which we are hypothetically operating. 
    \item For each interface element symbol $i \in I$,  $[[i]]$ maps $i$ to a unique meaningful segment of the interface such as a button, a list item, a form, a paragraph of text or a scrollbar.
    
    \item $[[I]]$ is the set $\{ [[i]] ~| \ i \in I \} $, so that $[[I]]$ refers to a complete interface or window.
    \item $[[\Omega]]$ is the set of all values which are possible interactor inputs: in the case of a computerized interactor, this would be a set of all possible data artifacts such as strings, booleans, integers or images that would be recognised by the interactor's own program language as program arguments.
    % In the case of an end-user, this would be more like a description of all physical procedures available to the end user to input a given value. 

    \item $[[F]]$ is the set of all valid action functions in the interactor language, so that each functional symbol $f \in F$ is mapped to a \emph{valid} action function of matching arity, with arguments interpreted pointwise according to the interpretation function.
    
    % $$[[f]] :[[S]] \times \bigcup \{[[\Omega]], [[I_1]],\ldots,[[I_n]]\} \to [[S]]$$
    % with computational state and arguments interpreted with respect to the context. 
\end{itemize}

For notational simplicity, we will henceforth conflate environmental elements with their interpretations (for example, $I$ for $[[I]]$) and suppress the state space argument to action functions. 

\end{mydef}

\begin{mydef}
An interface element $i$ is \textbf{positionally realised} if there exists a coordinate pair $bb(i) = ((x_0, y_0), (x_1, y_1))$ defining the bounding box associated with the instantiation of $i$ in the graphical user interface.

\begin{figure}[htbp]
    \centering
    \resizebox{0.5\columnwidth}{!}{
    \begin{tabular}{|c|}
    \hline
        \newline \\
        $ I_1 $ \\ 
        \includegraphics[width=7cm, trim={1cm 13cm 17cm 1cm}, clip]{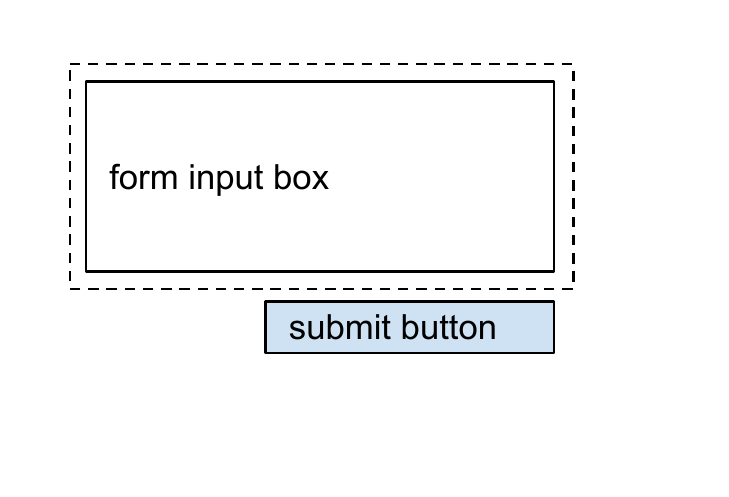}\\
        \hline
    \end{tabular}
    }
    \caption{Interface element $i_{1}^{(1)}$ with a positional realisation visualised as a bounding box.}
    \label{bounding_box_example}
\end{figure}

\end{mydef}

% \noindent \textbf{Definition (Positional realisation):} A function $f$ or and argument $arg$ is positionally realised (denoted by $f^*$) if there exists a coordinate pair (x,y) (where x in N and y in N) associated with the instantiation of $f$.
% \newline

%TODO: delete 
% An environment with all of its interface elements positionally realised with respect to a given desktop configuration \hl{TODO}

% \noindent \textbf{Definition (Resolvable naming references): } A function $f*$ or an argument $arg*$ are resolvable if there is a translation function $t$ such that:
% \newline

% \noindent $t : (x_0, y_0, x_1, y_1) \rightarrow \Gamma$ , $\forall f^*, arg* $ defined in e.
% \newline

% \noindent or defined alternatively: $t : Image \rightarrow \Gamma$ , $\forall f^*, arg* $ defined in e.
% \newline

% \noindent where $\Gamma \subset String$.
% \newline

A soft typing system may be implemented if one introduces an \emph{environmental vocabulary}, by which descriptive types may be assigned to both interface elements and input values. 

\begin{mydef}
An \textbf{environment vocabulary}  \mbox{$T \subset  \{`a\textrm',
\ldots,`z\textrm',`\ \ \textrm' \}^*$} is a set of descriptive strings strings. Define $type: \bigcup\{I_1, \ldots, I_n,  \Omega\} \to T$ to be a descriptor function for interface elements and values which may serve as action function arguments.
\end{mydef}

\begin{example}
In figure \ref{two_interface_example}, one may choose to define an environment vocabulary such that $type(i_1) = $ `button'. 
\end{example}

A descriptive vocabulary (or soft typing system) is not strictly necessary, but may serve for greater interpretability and potentially be used to introduce constraints on function arguments. If the system designer wishes, one may similarly introduce an \emph{action vocabulary} to which action functions may be mapped.

The purpose of the formalization in this section has been to formally define a \emph{process} in the context of IPA. The following definitions complete this goal and allows us to define the core tasks of IPA and to suggest evaluation metrics.

\begin{mydef}
A process $p$ (with respect to a interpreted environment $e$) is an ordered sequence of action functions 

\begin{align*}
    p = \big( & f_0(arg_0^{(0)}, \ldots, arg_{n_0}^{(0)}), \\
    & f_1(arg_0^{(1)}, \ldots, arg_{n_1}^{(1)}), \\
    & \vdots \\
    & f_m(arg_0^{(m)}, \ldots, arg_{n_m}^{(m)})\big) 
\end{align*}

with valid action functions and arguments interpreted (and valid) in the specified environment. 
\end{mydef}

\begin{mydef}
In the case that an environment $e$ is instantiated (or \emph{realised}) with respect to a \emph{computational} interactor, its programming language is referred to as an \textbf{IPA Realisation Language}.
\end{mydef}

\begin{mydef}
An \textbf{IPA program} is a process \emph{p} in an environment $\emph{e}$ which is instantiated with respect to an IPA Realisation Language. 
\end{mydef}

Although the definition of a process is purposefully environment agnostic, for the purpose of the tasks described in the next section we specifically assume process to mean a process which is an \emph{IPA program}. In particular, the tasks named demo2process and text2process require the output to be an IPA process with respect to some IPA realisation language.

\section{IPA Modalities \& Tasks}

\begin{figure}[h]
    \centering
    \includegraphics[width=0.3\textwidth]{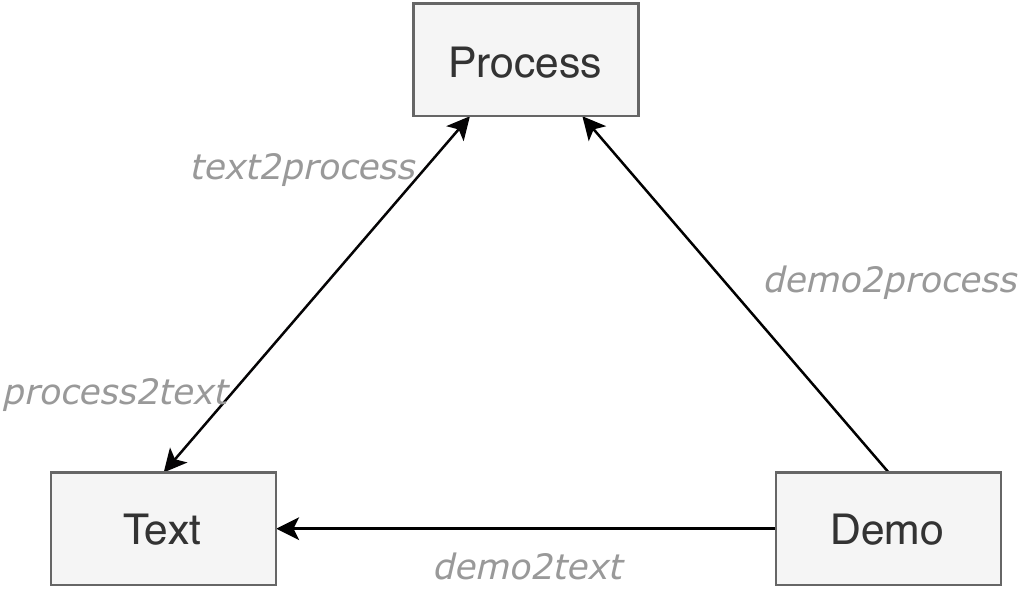}
    \caption{IPA tasks}
    \label{fig:triangle}
\end{figure}

\begin{figure*}[ht]
    \centering
    \includegraphics[width=0.7\textwidth]{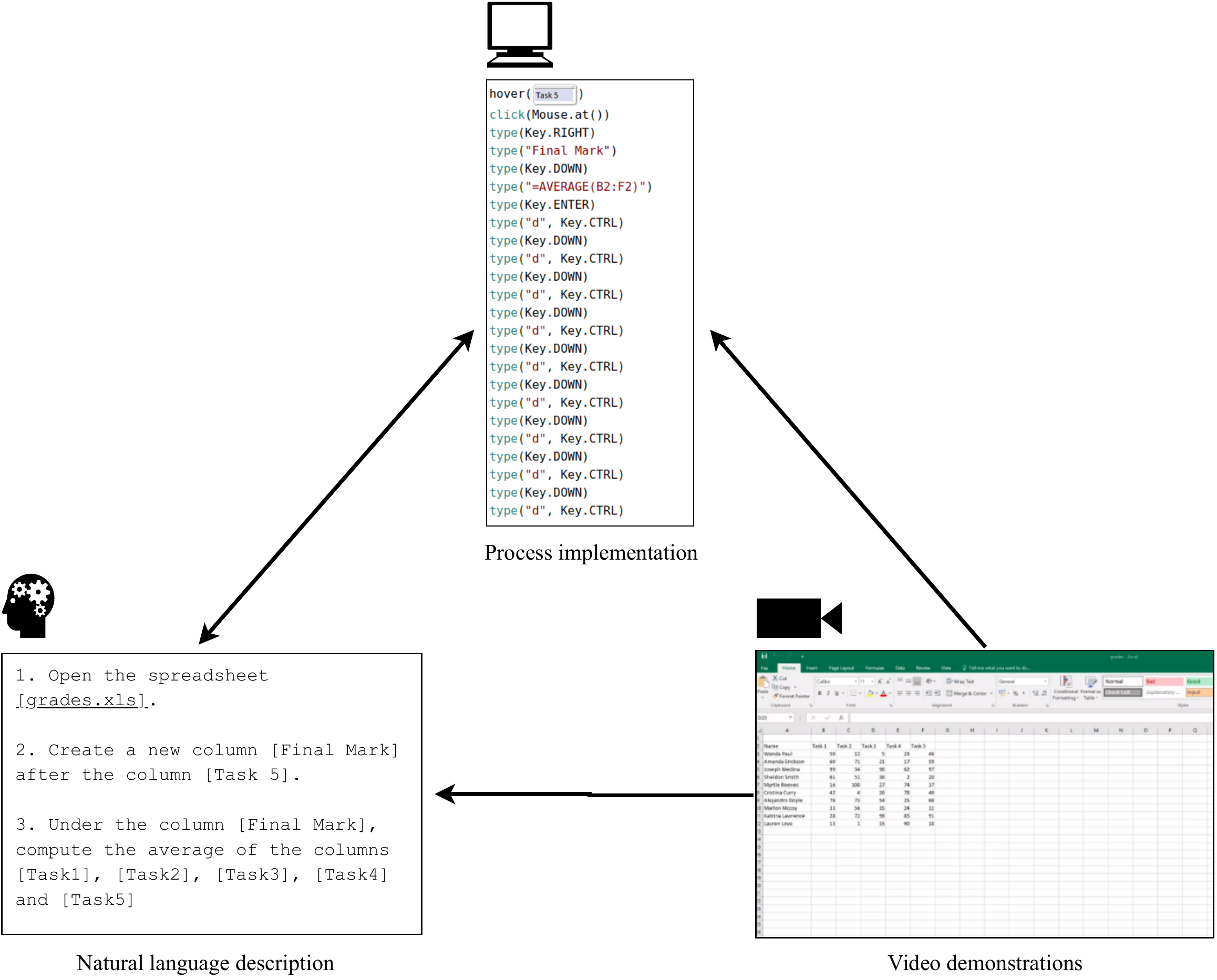}
    \caption{Example of IPA tasks}
    \label{fig:triangle_example}
\end{figure*}

Intelligent Process Automation can be defined as a combination of four different tasks: \textit{demo2text}, \textit{demo2process}, \textit{text2process} and \textit{process2text}, as shown in Figure~\ref{fig:triangle}. In this section we will describe each one of the tasks, using a motivational example presented in Figure~\ref{fig:triangle_example}.

With respect to our formalization, these tasks may be understood as the exercise of defining a mapping between the three different environment interpretations which refer to the same desktop configuration interpreting the \emph{interface} set \emph{I} but with action functions and arguments instantiated with respect to three different interactor ``languages": the recorded activity of a human (which constitutes a \emph{demo}), any IPA Realisation Language (which defines the target \emph{process}) and human natural language (instructional \emph{text}).

\subsection{demo2process (D2P)}

At the center of IPA is the task of transforming a demonstration to a program that is capable of re-executing the same process demonstrated by the user. We assume that all tasks are result-oriented, i.e., the user wants to achieve a specific result by executing the task. The results can be specific objectives, such as creating documents with specific values, or a set of actions, such as sending an e-mail to a client inquiring about the payment of a service.

Returning to the example in Figure~\ref{fig:triangle_example}, the resulting process is implemented using the Sikuli GUI Automation Tool. Running the Sikuli program, we obtain the same final spreadsheet as in in the demonstration.

\subsection{demo2text (D2T)}

The \textit{demo2text} task requires the conversion from a demonstration of a user executing an activity to a natural language description of the executed activity. In the example, the demonstration is a screen-recording of a user's desktop, divided into different time segments, where each segment has a corresponding natural language description of the actions. The natural language description should be a set of imperative sentences, where each sentence describes a single action that should be taken in order to replicate the demonstration.

The generation of natural language descriptions allows users to understand better the process that is being demonstrated, especially for users that are not executing the action, allowing new users to quickly learn how to execute the process themselves without external interference.

In Figure~\ref{fig:triangle_example}, we present the conversion from a video demonstrating a user manipulating a spreadsheet with student's grades to the corresponding natural language description of the task.

Demo2text is similar to the task of automated video captioning. However, video captioning is the task of describing general videos using natural language~\cite{pan2017video}, while demo2text focuses on the description of Desktop actions (usually materialised in video form) with imperative sentences.

\subsection{text2process (T2P) and process2text (P2T)}

Instead of having the end-user instance workflow  demonstration, it is possible to target a text description of the workflow. 
The text2process task is particularly useful when the user wants to generate the process from the description of the activity. The task is similar to end-user programming and semantic parsing, giving that users can convert from natural language to a program that implements a process~\cite{sales2017semeval} and Program Synthesis, since the final goal is generating a program. Given a set of imperative sentences, we want to generate a program that executes the described actions.

Similarly, the user might also want to generate the NL description of a process that is already implemented, especially in cases where the process becomes too complex to be analysed by humans.

\section{Related Work}

Intelligent process automation aims to replicate and improve, iteratively, activities carried out by humans. Comparing the performance of different IPA techniques requires well-defined tasks and metrics that reflect the relevant aspects when automating a process. In this section, we present the benchmarks present in similar research areas. 

Even though IPA applications, such as software intelligence and RPA, have particular research questions~\cite{aalst2018robotic}, there are only a few published datasets designed to help addressing this demand. 

One conspicuously related dataset is the Mini World of Bits (MiniWoB)~\cite{shi2017world}, a benchmark of 100 reinforcement learning environments containing many of the characteristics of live web tasks, created in a controlled context. Each MiniWoB environment is an HTML page with a resolution of  210x160 pixels, and a natural language task description, such as ``\textit{Click on the `Next' button.}''. The environment provides a precise evaluation metric, rewarding simulated behaviour, with rewards ranging from -1.0 (failure) to 1.0 (success), according to the results of each action, i.e., if the action shifts the current state to a state closer to the environment goal or not. 

The same work~\cite{shi2017world} also proposes FormWoB, which consists of four web tasks based on real flight booking websites, and QAWoB, which approaches web tasks as question answering, soliciting questions from crowd workers. Even though MiniWoB provides an essential baseline for IPA, it is still a synthetic dataset, not reflecting real user applications. It has a specific screen size, a limited number of applications and user actions, creating a controlled and closed environment. 

An example of a real (non-synthetic) dataset for software intelligence is the PhotoShop Operation Video (PSOV) Dataset~\cite{cheng2018learning}, containing videos and dense command annotations for Photoshop Software, with  74 hours of videos and 29,204 labelled commands. Despite PSOV presenting real-world use cases, it is still limited to one specific software, i.e., it is not generalisable to other applications.

Comparable benchmarks can be found in the research area of video captioning. \cite{rohrbach2015dataset}  presents the MPII Movie Description dataset (MPII-MD) which contains transcribed and aligned audio descriptions and script data sentences of a set of 55 movies of diverse genres. \cite{torabi2015using} introduces a dataset including 84.6 hours of paired video/sentences from 92 DVDs, with high-quality natural language phrases describing the visual content in a given segment of time.

A process can be seen as a formal representation of a task; therefore, there is an evident alignment between IPA and semantic parsing (SP). Unlike IPA, several benchmarks are available for SP, from which we can obtain insights. 

Most of these benchmarks focus on evaluating the conversion from a natural language specification to a program written in a specific programming language. WikiSQL~\cite{zhong2017seq2sql} is a collection of questions, corresponding SQL queries, and SQL tables. NL2Bash~\cite{lin2018nl2bash} is a corpus with frequently used
Bash commands with its respective natural language description. Other applications include converting from a visual object to code~\cite{ling2016latent,beltramelli2018pix2code} and from source code to Pseudo-code in different languages~\cite{oda2015learning}.

Process mining is also another closely related research area; it aims to discover, monitor and improve real processes, extracting knowledge from event logs~\cite{van2011process}. Unlike IPA, it does not have its main focus on automation, but on finding answers for domain-specific questions, such as analysing patient treatment procedures~\cite{bose2011analysis}, and discovering the roles of the people involved in the various stages of a specific process~\cite{van2015bpi}.

% \subsection{Related metrics - \hl{TODO} have better section title - Maybe we work on this after we have our metrics...}

% IPA output two possible outputs: texts and processes (or programs). In this section, we look at different existing methods of evaluating these two different outputs. 

%  BLEU~\cite{papineni2002bleu} has been initially proposed as a metric for automatic evaluation of machine translation; however,  different research fields have been applying this metric, including code generation. \cite{ling2016latent} evaluates code generation at the token level using BLEU; however, the authors realise that the generated source code can be correct without matching the reference. The work presents a possible solution to this issue, by verifying not only the generated coded, but also comparing the generated output with the output annotated (e.g. query results). 

% \section{Metrics for the Evaluation of IPA}

% Other relevant “metrics” or evaluation schemes for the problem

% Program Synthesis

% Very little consistency, not normalised

% NL to Process Models

% Mention how they evaluate

% Video captioning

\section{Evaluation Metrics}

\subsection{D2P \& T2P}
% TODO: Maybe make this distinction in the camera-ready iteration? 
% Three main categories of D2P \& T2P metrics are described: (i) retrospective IPA program, (ii) abstractive IPA program and (iii) output driven.

%\subsubsection{Metrics for IPA Programs}
 This section concentrates on a weighted quantification of the correctness and completeness of the generated IPA program output. Correctness and completeness are defined against a gold reference IPA program which was produced by one or more human programmers.

Given a set $\Pi = \{ p_1, \ldots p_k\} $ of generated IPA programs where each
$ p_j = f_0(arg_0, ... ,arg_n) \circ .... \circ f_m(arg_0, ... ,arg_n)$ and a corresponding gold standard $\Pi^{\star}$, we define a set of metrics of varying granularity that can be seen as approximate measures of program correctness. 

We start with the following error functions:
\begin{itemize}
    \item \textbf{Strict Error:}
    
    $E_{strict}(p, p^{\star}) = 
    \begin{cases}
    0 &\mbox{if} ~p_j = p_j^{\star} \\
    1 &\mbox{otherwise}
    \end{cases}$
    
    $MAE_{strict}(\Pi) = $
    $$\sum_{p_j \in \Pi}{\frac{E_{strict}(~p_j,p_j^{\star})}{|\Pi|}} $$

    \item \textbf{Predicate / Argument Sensitive Error:}
    
        $E_{sensitive}(p, p^{\star}) = $
        $$\sum_{\stackrel{f 
        \in \{f_0, \ldots, f_m \},}{arg \in  \{arg_0, \ldots, arg_n\}}}
        {\frac{E_{arg}(arg,arg^{\star})+E_{pred}(f,f^{\star})}{|\Pi|}} $$

    $E_{pred}(f, f^{\star})= 
    \begin{cases}
    0 &\mbox{if} ~f = f^{\star} \\
    1 &\mbox{otherwise}
    \end{cases} $
    
    $E_{arg}(arg, arg^{\star})= 
    \begin{cases}
    E_{image\_arg} &\mbox{if $arg$ is an image}  \\
    E_{symb\_arg} &\mbox{if $arg$ is a symbolic}\\
    & \mbox{value}
    \end{cases} $

\end{itemize}

% \hl{TODO: Maybe leave this distinction for next iteration?} This category concentrates on a weighted quantification of the correctness and completeness of the generated retrospective IPA program output. Correctness and completeness are defined against a gold reference retrospective program which was produced by one or more human programmers.
% \newline

% % Define precision, recall and F1-score here. 

% Given a set of retrospective IPA programs generated $\Pi_i = p_0(arg_0, ... ,arg_k) \circ .... \circ p_j(arg_0, ... ,arg_n)$ and a corresponding gold-standard $\Pi^*_i$.
% \newline

For an image argument, the intersection over union (IoU) is used to define $E_{image\_arg}$. Given the bounding box of the corresponding interface element ($bb(i(arg))$) and the gold standard bounding box ($bb^{\star}(i(arg))$):

$$IoU = \frac{\mbox{area\ of\ }( bb(i(arg)) \cap bb^{\star}(i(arg)))}{\mbox{area\ of\ } ( bb(i(arg)) \cup bb^{\star}(i(arg)))}$$

\noindent where $E_{image\_arg}(arg)= 0$ if $IoU > 0.5$ and $E_{image\_arg}(arg) = 0$ otherwise. 

IoU assumes that Image$_{sys}$ can be registered within the gs reference Screenshot$_{gs}$. 

An alternative way to define $E_{image\_arg}$ is by directly comparing the two argument images using means squared error (MSE) or structural similarity index (SSIM)~\cite{wang2004image}:
\newline
$$MSE = \frac{1}{mn} \sum \limits_{i=0}^{m-1} \sum \limits_{j=0}^{n-1} [I(i,j) - K(i,j)]^2$$

$$SSIM(x,y) = \frac{(2 \mu_x \mu_y + c_1)(2 \sigma_{xy} + c_2)}{(\mu^2_x + \mu^2_y + c_1)(\sigma^2_x + \sigma^2_y + c_2)}$$
\newline
\newline
The previous measures do not take into account the order of the program statements. In order to capture the sequential nature of an IPA program we include a sequence-based metric which is based on the longest common subsequence (LCS) function. Given two sequences $X = (x_1, x_2,  \ldots,  x_m)$ and $Y=(y_1, y_2, \ldots,  y_n)$, and given that the prefixes of $X$ are $X_{1,2,...,m}$ and the prefixes of $Y$ are $Y_{1,2,...,n}$,  LCS is defined as:
\newline 
$$
\mathit{LCS}(X_i,Y_j)=\begin{cases}
  \emptyset \\
  \qquad \quad (\mbox{if }i=0\mbox{ or }j=0) \vspace{1.5mm} \\
  \mathit{LCS}(X_{i-1},Y_{j-1})\text{ } \hat{}\text{ } x_i  \\
  \qquad \quad (\mbox{if }i,j>0\mbox{ and }x_i=y_j) \vspace{1.5mm} \\
  \operatorname{\max}\{\mathit{LCS}(X_i,Y_{j-1}),\mathit{LCS}(X_{i-1},Y_j)\}  \\
  \qquad \quad (\mbox{if }i,j>0\mbox{ and }x_i\ne y_j).
\end{cases}
$$
\newline
In order to compute the maximal program fragment generated we encode the programs $\Pi^{\star}$ and $\Pi$ as a sequence of unique symbols $s_i$ from a hash table derived from the statements of $\Pi^{\star} \cup \Pi$. The maximum program overlap (MPO) is defined as:
\newline
$ MPO(\Pi, \Pi^{\star}) = \frac{lcs(S,S^{\star})}{|S|}$
\newline
\subsection{P2T \& D2T}

Evaluating generated text is a requisite for areas related to Natural Language Generation, such as Machine Translation, Automatic Summarisation, Image/Video Captioning, and Document Similarity. BLEU has been widely applied and accepted as an evaluation metric for the tasks in the mentioned areas~\cite{gatt2018survey}. 

Similarly, we apply BLEU~\cite{papineni2002bleu} to evaluate text generated from demonstrations and processes. Our goal is to automatically evaluate for a demonstration $D_i$ or a process $P_i$ how well a candidate generated sentence $c_i$ matches the set of of demonstration/process descriptions $S_i = \{ s_{i1}, ..., s_{im} \}$. BLEU computes the n-gram overlap between the generated text description and the reference description. BLEU score depends on two different other factors: modified n-gram precision and brevity penalty.

Modified precision score computes the fraction of words matched between candidate descriptions and reference descriptions in the entire test corpus. Differently from precision, it clips the n-grams when it matches with the reference, avoiding high-precision results obtained with only repetitions of correct words. Modified n-gram precision is computed as follows:

\[
 p_n = \frac{\sum \limits_{\mathcal{C} \in \{ Candidates\}} \sum \limits_{ n \mhyphen gram \in \mathcal{C}} Count_{clip}(n \mhyphen gram)}{\sum \limits_{\mathcal{C} \in \{ Candidates\} } \sum \limits_{n \mhyphen gram \in \mathcal{C}} Count(n \mhyphen gram)}
\]

The brevity penalty factor penalizes candidates descriptions, $c$, shorter than the reference descriptions, $r$, computed as follows:

\[
 \text{BP} = 
  \begin{cases} 
   1 & \text{if } c > r \\
   e^{(1-r/c)}       & \text{if } c \leq r
  \end{cases}
\]

Finally, the BLEU score is calculated as
\newline
\[
 \text{BLEU} = BP \cdot \exp{(\sum\limits_{n=1}^{N} w_n \log p_n)} 
\]

where $w_n$ are positive weights summing to one and N is the size of the n-grams.

\section{Real World of Bits Benchmark}
To the best of our knowledge, there are no datasets available that can be used as a benchmark for all tasks defined in this work. Therefore, we create a benchmark for evaluation of IPA approaches. Inspired by the MiniWoB~\cite{shi2017world}, we name our dataset  \textit{Real World of Bits} (\textbf{RealWoB}). In this section we describe how we generated our benchmark.

RealWoB contains 100 different entries, where each entry is related to one specific task (e.g.,  searching for a flight) and it contains:
\begin{itemize}
\item A screen recording (video) of a user performing the task; 
\item A natural language summarisation of the task;
\item A natural language step-by-step description of the task;
\item A program that is capable of re-executing the task, written in Sikuli GUI Automation Tool and TagUI\footnote{https://github.com/kelaberetiv/TagUI}.
\end{itemize}

Each video has an average of 56 seconds and is divided into time intervals, where each interval has a corresponding natural language description. 

There are two types of natural language description for the task: a more general idea of the task being performed, i.e., a summarisation,  and a detailed step-by-step description. The summarised text is used as a guide for the annotator, in order to generate the video, description and program. The detailed description is written using imperative sentences, such as \textit{``Click on the button `Send"}. 

\begin{figure*}[h]
    \centering
    \includegraphics[width=0.6\textwidth]{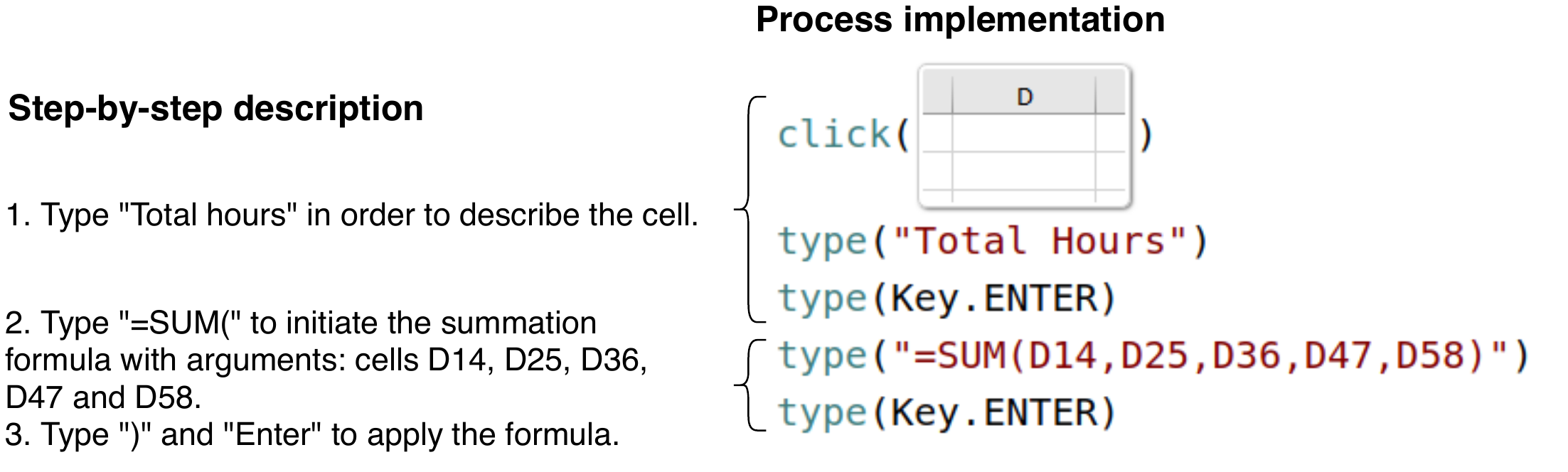}
    \caption{NL description and respective Sikuli process implementation.}
    \label{fig:nl_to_sikuli}
\end{figure*}

RealWoB also contains a program for every task. The programs were implemented based on the videos and the natural language step-by-step. Every sentence in natural language corresponds to one or more commands in the program/process. Figure~\ref{fig:nl_to_sikuli} presents an example of pair of natural language step-by-step description process implementation in Sikuli.

The 100 tasks are divided into 10 categories:
\begin{itemize}
\item Spreadsheet use: 10 tasks
\item Spreadsheet and browser use - simple: 10 tasks 
\item Spreadsheet and browser use - elaborate: 10 tasks
\item Webmail suite use: 10 tasks
\item Spreadsheet and Webmail suite use: 10 tasks
\item Webmail suite and browser use: 10 tasks
\item  Browser, spreadsheet and Webmail suite use: 10 tasks
\item Browser (social media) use: 10 tasks
\item Browser (social media) and spreadsheet use: 10 tasks
\item Random selection of previous tasks executed in a different operating system: 10 tasks
\end{itemize}

The selected group of tasks/computer applications are common for office workers, with a high number of potential end users. We identified the tasks as suitable for automation, where none or little human input is required. 

The automation of tasks using webmail suites and web browsing is particularly hard due to the dynamic state of the web. For tasks involving receiving/sending e-mails, we consider that all e-mails follow a pre-defined template. For the ones that involve browsing the web, we cache a version of the used webpage and assume it as the current state of the page.

For illustration purposes, Table~\ref{tab:task_examples} presents one example of each task.

% \usepackage{booktabs}
% \usepackage{graphicx}
\begin{table*}[ht!]
\centering
\caption{Examples of tasks in dataset RealWoB.}
\label{tab:task_examples}
\resizebox{\textwidth}{!}{%
\begin{tabular}{@{}ll@{}}
\toprule
\multicolumn{1}{c}{Category}         & \multicolumn{1}{c}{Task Example (Summary)}                                                                                                                                                                                                                                                                                            \\ \midrule
Spreadsheet                          & \begin{tabular}[c]{@{}l@{}}Open a spreadsheet \textless{}file location\textgreater  containing columns for names of students and marks for five assignments. \\ Create a new column containing the mean of the marks for each student.\\ \end{tabular}                                      \\                                              \\
Spreadsheet + browser                & \begin{tabular}[c]{@{}l@{}}Open the website: https://veggiedesserts.co.uk/vegan-carrot-cake/\\ Find the ingredients for the recipe on the page.\\ Open a spreadsheet \textless{}file location\textgreater containing a column for ingredients and a column for quantity.\\ Copy the recipe details into the spreadsheet.\end{tabular} \\
\\
Spreadsheet + browser                & \begin{tabular}[c]{@{}l@{}}Open a spreadsheet \textless{}file location\textgreater containing a list of products.\\ Search each item on https://www.idealo.co.uk/\\ Add the lowest price in the column next to each product name.\end{tabular}                                                                                        \\
\\
Webmail                              & \begin{tabular}[c]{@{}l@{}}Open outlook.\\ Write an email to \textless{}hidden\textgreater requesting to book a flight from Manchester to Tokyo on the 01/08/2019.\end{tabular}                                                                                                                                                       \\
\\
Spreadsheet + Webmail                & \begin{tabular}[c]{@{}l@{}}Open a spreadsheet containing names of people, places (origin and destination) and dates (in a spreadsheet),\\ Open outlook.\\ Send an email to \textless{}hidden\textgreater requesting a flight booking for each row in the spreadsheet (one e-mail for each).\end{tabular}                              \\
\\
Webmail + browser                    & \begin{tabular}[c]{@{}l@{}}Open outlook.\\ Read an email requesting the booking of a flight. \\ Use https://www.skyscanner.net/ to verify the price of the flight.\end{tabular}                                                                                                                                                       \\
\\
Browser + spreadsheet + webmail      & \begin{tabular}[c]{@{}l@{}}Open outlook.\\ Read one email requesting the booking of a flight.\\ Use https://www.skyscanner.net/ to verify the price of the flight. \\ Add the name of the person and the cheapest flight to a (pre-created) spreadsheet.\end{tabular}                                                                 \\
\\
Browser (social media)               & \begin{tabular}[c]{@{}l@{}}Go to https://www.imdb.com/chart/top?ref\_=nv\_mv\_250.\\ Search for the trailers of the top 5 movies of all time on Youtube using the name of the movie and the word “trailer”.\end{tabular}                                                                                                              \\
\\
Browser (social media) + spreadsheet & \begin{tabular}[c]{@{}l@{}}Find the current top 10 hashtags on Twitter.\\ Add the hashtags it to a pre-created spreadsheet.\end{tabular}                                                                                           \\                                                                                                   \\
Random task - Different OS           & Any of the above tasks executed using the OS Ubuntu.                                                                                                                                                                                                                                                                                  \\ \bottomrule
\end{tabular}%
}
\end{table*}

RealWoB was annotated by four different annotators, A, B, C and D, where C is a specialist in IPA. It was constructed with the following steps:

\begin{enumerate}
\item We defined 90 different tasks based on investigating everyday office worker tasks.
\item Annotator A recorded 90 videos, running the tasks on Windows 10.
\item Annotator B wrote a summary of the task being performed, a step by step description (for every video segment) and a program capable of executing the task automatically.
\item Annotator C verified and corrected the videos and the annotations.
\item From the 90 different tasks, we chose ten random tasks.
\item Annotator B recorded the ten tasks using Ubuntu 16.04.
\item Annotator D wrote a summary of the task being performed, a step by step description (for every video segment) and a program capable of executing the task automatically.
\item  Annotator C verified and corrected the videos and the annotations.
    
\end{enumerate}

\section{Conclusion}

In this work, we present a formalisation for IPA using a bottom-up approach, defining several composing blocks of an IPA program. We also define the modalities and tasks that are part of IPA, demo2text, demo2process, text2process and process2text, and we present how these tasks relate to similar research areas. 

With the formalisation, it was also possible to define appropriate metrics for each one of the tasks, evaluating the resulting process or text. We also describe how we built a dataset that can be used as a benchmark for the IPA tasks.

We envision that the research area in IPA will see significant progress in the next few years, following the increased use of RPA tools. The formalisation present in this work is a starting step towards building complex IPA systems.

While we have provided the formalisation, evaluation techniques and methodology for designing a benchmark, there is an important next step in this research: joining together these contributions and creating a baseline for each one of the IPA tasks.

\subsection{Acknowledgements}
The authors would like to thank Jacques Cali and the Blue Prism team for their support of this project. We also thank the anonymous reviewers for the valuable feedback.

\bibliography{ref}
\bibliographystyle{aaai}

\appendix

\end{document}